\documentclass{aa}
\usepackage{graphicx}
\usepackage{latexsym}
\usepackage{amssymb}
\usepackage{amsmath}
\usepackage[]{natbib}
\bibpunct{(}{)}{;}{a}{}{,}
\newcommand{\etal}{{\it et al.}}

\def\sec{\ifmmode{^{\prime\prime}}\else{$^{\prime\prime}$}\fi}
\def\min{\ifmmode{^{\prime}}\else{$^{\prime}$}\fi}
\def\deg{\ifmmode{^\circ}\else{$^\circ$}\fi}
\def\arcsec#1.#2 {\ifmmode {#1^{\prime\prime}\hskip-0.42em.
                  \hskip0.15em#2}
         \else {$#1^{\prime\prime}\hskip-0.42em.\hskip0.15em#2$}
         \fi}
\def\arcmin#1.#2 {\ifmmode {#1^{\hskip 0.05em\prime}\hskip-0.35em.
                  \hskip0.05em#2}
         \else {$#1^{\hskip 0.05em\prime}\hskip-0.35em.\hskip0.05em#2$}
         \fi}
\def\arcdeg#1.#2 {\ifmmode {#1\deg\hskip-0.42em.
                  \hskip0.10em#2}
         \else {$#1\deg\hskip-0.42em.\hskip0.10em#2$}
         \fi}

\begin{document}

\title{Relations between ISM tracers in galaxies}

\author{G. Galletta\inst{1}
        \and V. Casasola\inst{1,3}
          \and L. Piovan\inst{1,4}
           \and E. Merlin\inst{1}
            \and D. Bettoni\inst{2} }

\offprints{giuseppe.galletta@unipd.it}

\institute{Dipartimento di Astronomia, Universit\`a di Padova, Vicolo
               dell'Osservatorio 2, I-35122, Padova\\
             \email{giuseppe.galletta@unipd.it, viviana.casasola@unipd.it,
                     piovan@pd.astro.it, merlin@pd.astro.it}
         \and INAF - Osservatorio Astronomico di Padova, Vicolo dell
                'Osservatorio 5, I-35122, Padova\\
             \email{daniela.bettoni@oapd.inaf.it}
          \and
     Observatoire de Paris-LERMA, 61 Av. de l'Observatoire, F-75014 Paris
          \and
     Max-Planck-Institut f\"ur Astrophysik, Karl-Schwarzschild-Str. 1, Garching bei M\"unchen,
     Germany\\
      \email{piovan@mpa-garching.mpg.de}}
\date{Received ; Accepted 20 September 2006}

\titlerunning{ISM tracers}

\abstract
   {}
    {We study the relations existing between fluxes emitted at CO(1-0) line, 60
and 100 $\mu$m wavelengths, B and soft X-ray wavebands for galaxies of all
morphological types. The large set of data that we created allows to revisit
some of known relations existing between the different tracers of the Interstellar
Medium (ISM): the link between the FIR flux and the CO line emission, the relation 
between X-ray emission in non active galaxies and the blue or FIR luminosity.}
   {Using catalogues of galaxies and works presented in the literature,
we collected fluxes in FIR, 21 cm, CO(0-1) line and soft X-ray for two samples,
consisting of normal and interacting galaxies respectively.
Joining together these samples, we have data for a total of 2953 galaxies, not
all observed in the four above wavebands.}
   { All the relations found are discussed in the frame of the star formation
activity that is the link for most of them. We note that when an active star
formation is present, it may link the galaxy fluxes at almost all wavelengths,
from X to microwaves. On the contrary, in early-type galaxies where the current
star formation rate has faded out the X-FIR fluxes link disappears. This result
obtained for early-type galaxies is discussed and explained in detail in the
frame of a suitable theoretical model, obtained coupling
chemo-dynamical N-body simulations with a dusty spectrophotometric
code of population synthesis.}
   {}

\maketitle

\section{Introduction}
The observations of galaxies at various wavelengths, going from
radio to X-ray, allow to study the relationships existing between
the various phases of the interstellar gas, and between gas, dust
and stars. Some of these relations are already known since many
years, such as that between CO and far infrared (FIR)
luminosities \citep{mirabel,sanders,solomon,devereux}. Others,
connected with X-ray emission, have been studied more recently
\citep{padovani,david, ranalli}.

\begin{figure*}
\centering
\resizebox{17cm}{!}{\includegraphics[angle=0]{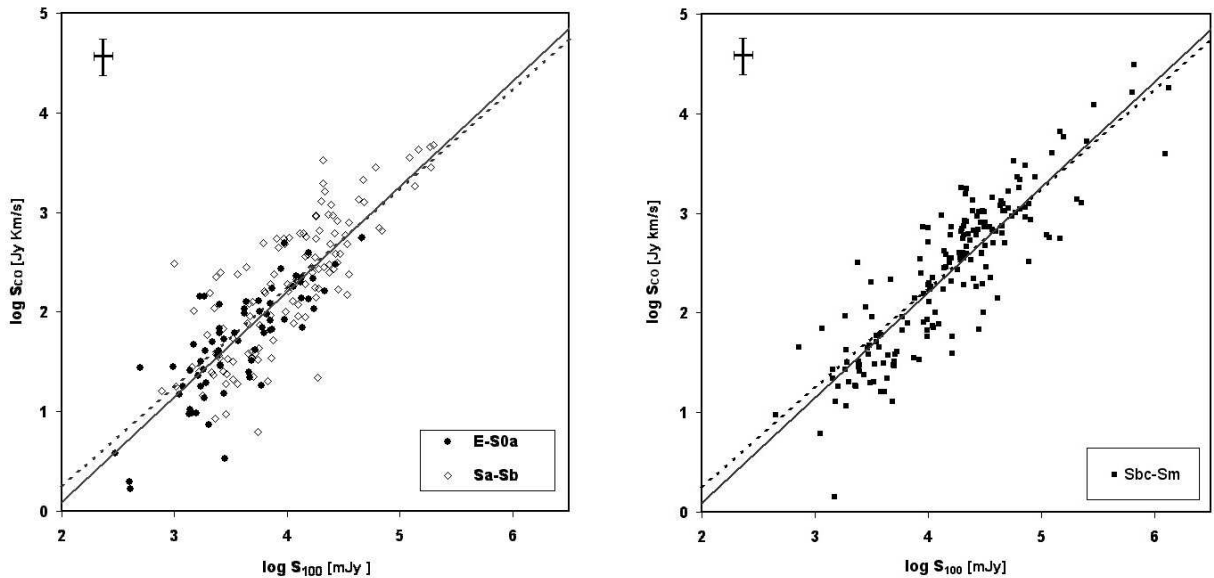}}
\caption{The relation between the flux from the CO(1-0) line and
that at 100 $\mu$m, plotted as a full line, with the representative
points of the single galaxies plotted with different symbols. The
crosses at the upper left of the figures represent the mean errors
for the fluxes in the whole sample. For comparison the relation
found by \citet{bregman} for a sample of early-type galaxies is
shown as a dotted line. The early-type galaxies and early spirals,
from E to Sb, are plotted in the left panel while the late spirals
are plotted in the right panel. } \label{CO_100}
\end{figure*}

At present, different tracers of the gas are known, such as
millimetric lines for the cold molecular gas, the 21 cm line for
atomic hydrogen at $\sim$100 K, IR bands for molecules at thousands
of degrees, UV lines and X-ray emission for hotter gas. The dust
distribution is traced also by FIR emission at 60 and 100 $\mu$m if
the grains are warm \citep{bregman} or at 170 $\mu$m, if they are
colder \citep{popescu}. The diffusion of large archives of
observations at the above wavelengths (except for molecular lines)
allowed in the last years the compilation of catalogues containing
a huge number of galaxies. Using these catalogues and the works
presented in the literature, we collected fluxes in FIR, 21 cm,
CO(0-1) line and soft X-ray for two wide samples of normal
\citep{normal} and interacting \citep{interacting} galaxies. Joining
together these samples, we have data for a total of 2953 galaxies,
not all observed in the four above wavebands. The fluxes measured
with the different tracers allow now a study of the
link existing between dust, gas and stars based on hundreds of
galaxies.

It is known that the fluxes emitted by a galaxy at very different
wavelengths may be linked together by means of the star formation
mechanism (see \citet{david,ranalli}).
For instance, the formation of massive stars generates the
heating of the dust clouds in which they are embedded, by absorption
of their UV radiation, and produces a re-emission of this energy in
the far infrared. This process links the current star formation rate
to the IR emission at 60 and 100 $\mu$m \citep{thronson}. The
ionizing radiation of stars may produce also the evaporation of the
molecular clouds. Inside these clouds, where the particle density is
great enough to produce a significant number of collisions between
H$_2$ and CO molecules, these latter are excited and produce
photons, but in optically thick regions. The warming by the UV
stellar light makes these regions less dense, making visible the CO
lines at their edge. Because of this mechanism, these lines are
considered tracers of the cold molecular hydrogen that does not emit
observable lines. The newly formed stars are also responsible of the
X-ray emission, produced by very massive stars, by core-collapse SN,
and by high mass X-ray binaries. According to the above described
mechanisms, we expect that galaxies with active star formation will
have a far infrared emission, but also CO and X-ray emissions induced by the
more massive stars, linked together by means of different relations.

When the star formation decreases or vanishes, the far infrared emission decreases
as well, but it may be fed by the stellar light absorbed and re-emitted
in the infrared by dust (cirrus), while low-mass X-ray binaries and Type I SN
contribute to the high energy galaxy spectrum. In addition, AGB stars,
surrounded by dust, and the cooling flows of the interstellar medium
ejected by supernovae may produce additional IR and X emission,
between each other.

To study the activity of the galaxies at different wavebands, we
collected data on galaxies starting from the original data of fluxes
at 60, 100 $\mu$m, CO(1-0) 2.6 mm and soft X-ray used to compile
our catalogues \citep{norm_cat, inter_cat}. The merging of the two
above catalogues produces 1764 known values of
far infrared fluxes (1837 have 100 $\mu$m flux), 391 soft X-ray fluxes and 434
values of the CO(1-0) line luminosity. We extracted from LEDA catalogue
\citep{leda} the values of the distance moduli, blue absolute magnitudes
and morphological classification for all of them.

Galaxies with evident sign of interactions or disturbed morphologies according
to the catalogues of \citet{arp,am,vv} are 1038. We shall refer to them as ``perturbed
galaxies''. The remaining 1915 galaxies that appear neither morphologically
nor dynamically perturbed are called ``normal galaxies''.
In our sample, we have 253 galaxies that have spectral classification of the
nucleus and 231 of these appear to host an AGN (Seyfert 1, 2 or transition type,
Seyfert 3 or Liners) according to the classifications of \citet{ho} and
 \citet{veron}. Most part of the remaining 2722 galaxies lacks of information
about nuclear spectrum or have spectra of HII regions (22 starburst spectra).
They are not included in any AGN catalogue and for this reason in the following discussion
we refer to them as ``non active galaxies" and to the others as ``active
galaxies". With all these data, we crossed the various tracers to understand
and revisit the main relations existing between X, FIR, CO and B luminosities.

\begin{figure*}
\resizebox{17cm}{!}{\includegraphics[angle=0]{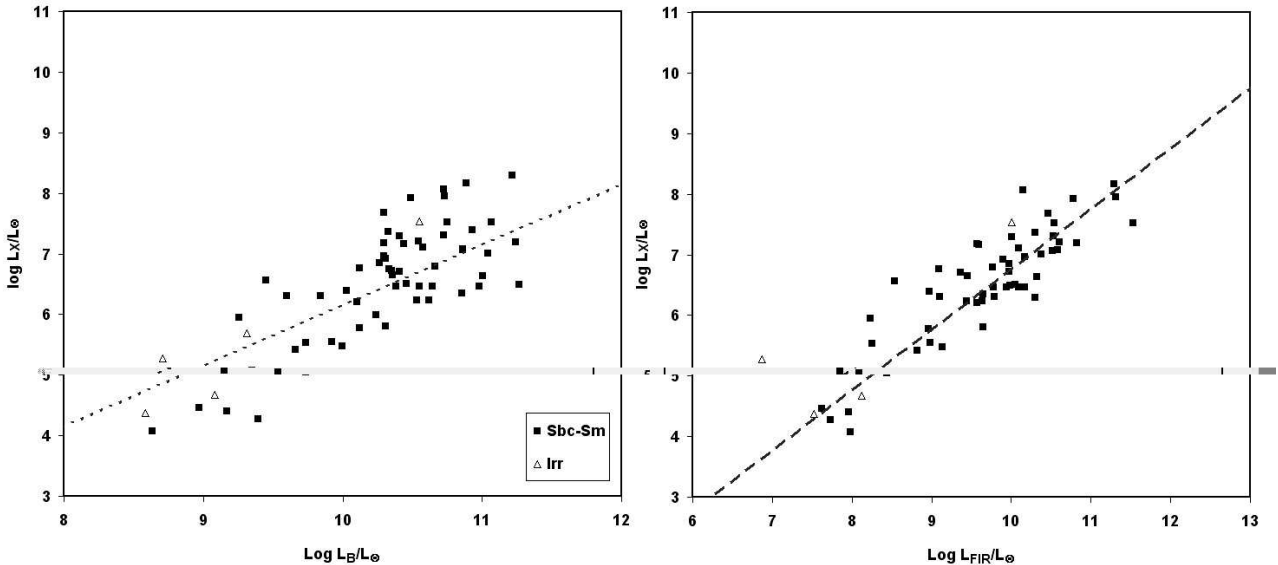}}
\caption{{\it Left:} The X-ray luminosity plotted vs. blue luminosity, in
solar units, for late-type, non-active galaxies. The relation corresponding
to emission caused by discrete sources, indicated in equation (\ref{XBlate})
is plotted as a dotted line.
{\it Right:} X-ray luminosity vs. FIR luminosity, in solar units
for the late-type, non-active galaxies. The linear relation indicated in equation
(\ref{XFIR}) is plotted as a dashed line.}
\label{X1}
\end{figure*}

\section{Cold gas and warm dust}

The relations existing between different cold components of the ISM such as
the molecular gas and the dust have been studied since many years
\citep{mirabel,solomon, bregman}. They find that the global galaxy luminosity
derived from CO(1-0) line is directly related with the flux at 100 $\mu$m.
With our large sample we can now test these relations using galaxies
of different morphological types and activity or interaction.

In Figure \ref{CO_100} we plotted the logarithm of the flux measured from CO(1-0)
line vs. the logarithm of the IRAS flux at 100 $\mu$m. In our plots, we have
193 galaxies with classification from E to Sb and 178 from Sbc to Sm. The
relation found by \citet{bregman} for a sample of early-type galaxies,
log S$_{CO}$=log S$_{100}$ - 1.76, is also
plotted as comparison, as a dotted line.

The relations are evident, with this wider sample of galaxies.
In these diagrams, active and non-active galaxies appear mixed together
without clear differences and have been plotted together. The same behaviour
appears for interacting and non interacting galaxies, that are not distinguished
in our plots.

For all the galaxy types, we find:
\begin{equation}
Log S_{CO}= 1.06\ Log S_{100} + 2.02
\label{eqCO_100}
\end{equation}
with a correlation coefficient of 0.74 and a r.m.s. of 0.37. In the above
formula, S$_{100}$ is in mJy and S$_{CO}$ is in Jy km/s.

Similar relations exist between the CO fluxes and the FIR magnitudes, defined as:
\begin{equation}
m_{FIR} = -2.5\ Log(2.58\ S_{60}+S_{100})+ 22.25
\label{def_mfir}
\end{equation}
where S$_{60}$ and S$_{100}$, the fluxes at 60 and 100 $\mu$m respectively,
are in mJy. We find for all the galaxy types:
\begin{equation}
Log S_{CO} = 0.41\ m_{FIR}+ 6.86
\label{CO_mfir}
\end{equation}
with a correlation coefficient of 0.69 and a r.m.s of 0.40.
The results are based on 179 early types and 170 late-type galaxies.
For their similarity with Figures \ref{CO_100} these relations are not plotted
in this paper.

We note that irregular galaxies are not fitted by these relations but
have a wide spread. In our sample there are just 10 galaxies and their
representative points have been not plotted in Figure \ref{CO_100}.

\section{X-ray component.}

\begin{figure*}
\resizebox{17cm}{!}{\includegraphics[angle=0]{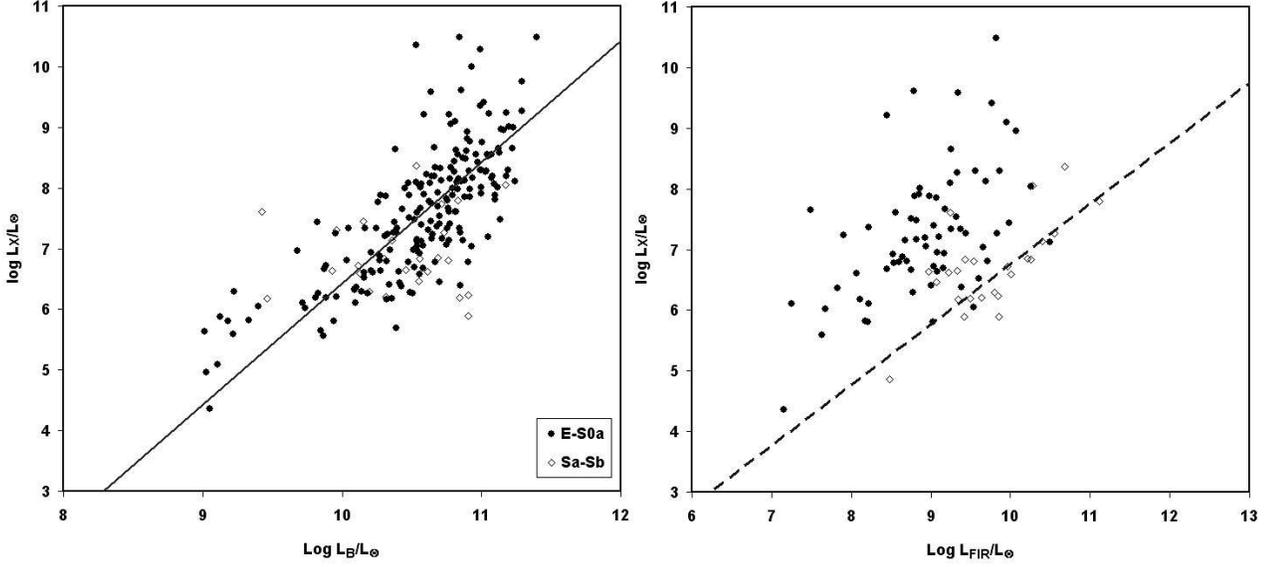}}
\caption{{\it Left:} The X-ray luminosity plotted vs. blue luminosity,
for early-type, non-active galaxies. The relation corresponding to emission
caused by diffuse gas, valid for early-type galaxies and indicated in equation
(\ref{XBearly}) is plotted as a full line.
{\it Right:} X-ray luminosity vs. FIR luminosity for the early-type galaxies.
The linear relation for late-type, non-active galaxies indicated in equation (\ref{XFIR})
is plotted as a dashed line.}
\label{X2}
\end{figure*}

We are interested to understand what relations exist between L$_X$, the X-ray
luminosity, and the other global galaxy properties. From the literature, it is
known the existence of a proportionality between L$_X$ produced by
discrete sources and L$_B$, the blue luminosity of the whole galaxy.
This relation has been studied by \citet{ciotti} and compared by \citet{beuing}
with soft X-ray fluxes measured by ROSAT satellite. It appears that late-type
galaxies have a global X-ray luminosity directly proportional to L$_B$, while
early-type systems are dominated by emission produced by hot diffuse gas and
their $L_X$ is proportional to the square power of the blue luminosity, as
discussed by \citet{beuing}. For this reason, the early and late-type
galaxies are discussed separately.

\subsection{Late-type galaxies}
With our data, the X-ray luminosity of galaxies with morphological type later
than Sb can be fitted by a linear relation as a function of L$_B$(dotted line
in Fig. \ref{X1}, left panel). The direct proportionality is expressed by the
equation:
\begin{equation}
Log L_X = Log L_B - 3.85
\label{XBlate}
\end{equation}
with a r.m.s. from observed data of $\sigma$=0.61 based on 63 galaxies. In this
formula and in the following, all the luminosities are expressed in solar units.

If, instead of the blue luminosity, we use the galaxy area $D^2_{kpc}$, calculated
from the apparent diameter measured at the 25 mag arcsec$^2$ isophote and
converted in kpc$^2$, we
discover that the relation is still present, but with a larger spread. It becomes:
\begin{equation}
Log L_X = Log D^2_{kpc} + 3.83
\end{equation}
($\sigma$=0.80) for a sample of 64 galaxies.

A relation similar to that of \citet{ciotti} has been found by some authors
\citep{padovani, david, ranalli}, but using 60 $\mu$m fluxes or FIR
luminosities. The values of $L_{FIR}$ are calculated using the formula:
\begin{equation}
Log L_{FIR}= 2.59+Log(2.58\ S_{60}+S_{100})+2 Log\ d
\end{equation}
where L$_{FIR}$ is in solar luminosities, fluxes are in mJy and the galaxy
distance d is in Mpc.

\begin{figure*}
\resizebox{17cm}{!}{\includegraphics[angle=0]{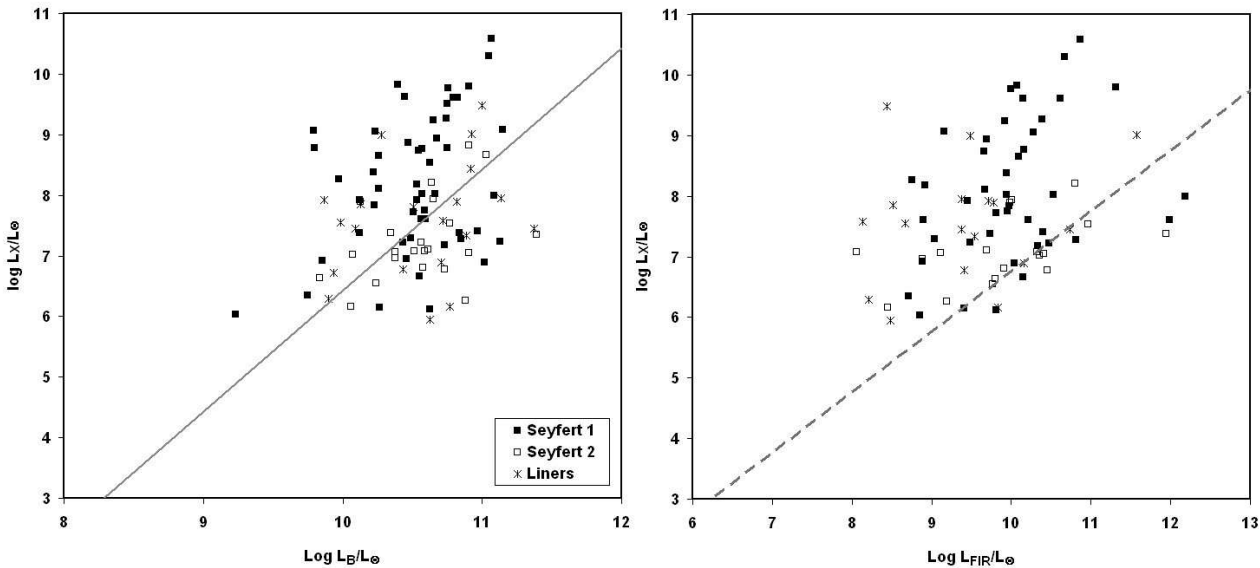}}
\caption{{\it Left:} The X-ray luminosity plotted vs. blue luminosity,
for active galaxies, separated according the type of activity. The relation corresponding to
emission caused by diffuse gas, valid for early-type galaxies and indicated in equation
(\ref{XBearly}) is plotted as a full line. {\it Right:} X-ray luminosity vs. FIR luminosity
for the active galaxies. The linear relation for late-type galaxies indicated in equation
(\ref{XFIR}) is plotted as a dashed line.}
\label{X3}
\end{figure*}

From our data it is possible to find a relation between L$_X$ and L$_{FIR}$
that fits the values of late-type galaxies. We found L$_X \propto$
L$_{FIR}^{0.90}$, similar to the L$_X \propto$ L$_{FIR}^{0.88}$ found by
\citet{ranalli} for fluxes between 0.5 and 2 keV and to L$_X \propto$
L$_{FIR}^{0.95}$ found by \citet{david} using fluxes between 0.5
and 4.5 keV. Forcing the relation to a linear proportionality between
L$_X$ and L$_{FIR}$ we find:
\begin{equation}
Log L_X = Log L_{FIR} - 3.18
\label{XFIR}
\end{equation}
with a $\sigma$ of 0.47, based on 147 galaxies.
This relation is plotted as a dashed line in the right panels of Figures from
\ref{X1} to \ref{X3}.

We note that the B and FIR luminosities are also connected in late-type
galaxies by means of a linear relation fitted by:
\begin{equation}
Log L_{FIR} = Log L_B - 0.38
\end{equation}
with a r.m.s.=0.5. This equation, inserted into the relation (\ref{XBlate})
gives:
\begin{equation}
Log L_X = Log L_{FIR} -3.47
\end{equation}
similar to the result of equation (\ref{XFIR}) and to that found by
\citet{ranalli}. This is an independent way to confirm our results
and to verify the existence of a global link between L$_{FIR}$, B
light and X-ray emission. The connection between B luminosity
or galaxy area and X or FIR luminosities will be discussed in
Section \ref{Discussion}.

\subsection{Early-type galaxies}
When the early-type galaxies are considered in the above described relations
involving X-ray emission, the correlations become less evident. Considering
soft X-ray and B luminosities, we find a relation:
\begin{equation}
Log L_X = 2\  Log L_B - 13.57
\label{XBearly}
\end{equation}
($\sigma$=0.73) based on 224 galaxies and plotted as full line in Fig. \ref{X2},
left panel. The above formula agrees with the expected relation for X-ray
emission coming from hot diffuse gas, as discussed by \citet{beuing}.

The relation still hold if $D^2_{kpc}$ (kpc$^2$) is used. It becomes:
\begin{equation}
Log L_X = 2 Log D^2_{kpc} + 1.51
\end{equation}
($\sigma$=0.85) for 226 early-type galaxies from E to Sb.

Many galaxies with high blue luminosity, indication of high masses and of a
recent star formation, lie quite far from the mean line, with a behaviour
different than that of late-type galaxies.

If the X-ray fluxes are compared with FIR luminosity, the disagreement with
the behaviour found in late-type galaxies is more evident. The plot L$_X$ vs.
L$_{FIR}$ for early-type galaxies shows the representative points of the
galaxies above the relation (\ref{XFIR}) for late-type galaxies
(Fig.\ref{X2}, right panel). To understand this apparent disagreement, we
should use a theoretical analysis of the far infrared emission, as explained 
in the next Section 4.

\subsection{Active galaxies}
Active galaxies (Seyfert 1, Seyfert 2 and Liners) have X-ray, B and FIR fluxes
that are not linked together. This happens because, to the emission mechanisms
stimulating the light emission at the different wavebands described for non
active galaxies, adds an X-ray emission coming from nucleus. In fact, the points
representative of these active galaxies are spread in the plot over the discrete
sources line and around the diffuse gas line (see Fig.\ref{X3}, left side).
In the L$_X$--L$_{FIR}$ diagram (Fig.\ref{X3}, right side) the spread is similar 
to that of early-type galaxies plotted in Fig.\ref{X2}, but we separately plotted 
the active galaxies because of the particular nature of their X-ray emission, 
due to the nuclear contribution.

\section{Modelling $L_{X}$, $L_{B}$ and $L_{FIR}$ of early-type galaxies}\label{Modelling}

To cast light on the nature of the relations observed
between $L_{X}$, \textbf{$L_{B}$} and $L_{FIR}$ for early-type
galaxies, one has to consider the various
components of a galaxy (stars, gas and dust) and to understand
their mutual interactions as far as the spectral energy
distribution (SED) is concerned. There are two basic schemes to
model the formation and evolution of early type galaxies: (1) the
semi-analytical models on which a great deal of our understanding of
the chemo-spectro-photometric properties is derived, and (2) the
N-Body Tree-SPH simulations which, in contrast, have been only
occasionally used to study spectro-photometric properties of
early type galaxies. In the following part of this section
we will proceed as follows. First we will analyse the drawbacks of
semi-analytical models, in particular dealing with the calculation
of the infrared emission of early-type galaxies. Second, we will
discuss how dynamical simulations and a dusty
spectrophotometric code, when mixed together allow to move a step
forward in the calculations of the SEDs properties. Third, we will
show in detail how our model has been built and the coupling between
dynamics and dusty population synthesis has been done.

\subsection{The semi-analytical models and their drawbacks}

The semi-analytical models approximate a galaxy to a point mass
system in which gas is turned into stars by means of suitable
recipes for star formation and heavy elements are produced by
stellar nucleosynthesis and stellar winds/explosions. The
standard evolutionary population synthesis technique (EPS) is
usually applied to derive the SED of the galaxy, with models
able to explain many global features of early type galaxies,
as amply described by many authors \citep{Arimoto87,Arimoto89,
Bressan94, Gibson97b, Tantalo96, Tantalo98}. There are three important and
problematic issues of these models to be discussed for our purposes.

First, to determine the age at which the galactic wind
sets \citep{Larson74,Larson75}, we need some hypothesis about Dark
and Baryonic Matter with their relative distributions, and about the
heating and cooling efficiency of the various mechanisms, to
properly evaluate the total gravitational potential well and to
describe the thermal history of the gas. In this scheme it comes
out that the galactic wind occurs typically for ages $t_{GW} < 1$ Gyr,
later in a massive early-type galaxy and much earlier in galaxies of lower mass
\citep{Arimoto87, Arimoto89, Bressan94, Gibson97b, Tantalo96, Tantalo98, Chiosi98}.
The maximum duration of star forming activity follows
therefore in these models the trend $\Delta t_{SF} \propto M_{G}$.
This trend of the SFH is, however, contrary to what required by the
observes trend of the $\alpha$-enhancement for early type galaxies,
which implies that the maximum duration of the star forming activity
should decrease when the galaxy mass increases $\left( \Delta t_{SF}
\propto M_{G}^{-1}\right)$ \citep[see][ for more details on the
enhancement in $\alpha$-elements and the SFH of early-type
galaxies]{Bressan96, Kuntschner00, Trager00a, Trager00b,
Tantalo04, Thomas05}.

Second, after the galactic wind phase, star formation does no longer
occur and the evolution is merely passive. However, AGB and RGB
stars continue to loose gas in amounts that are comparable
to those before the galactic wind \citep{Chiosi00}. What is
the fate of this gas? One may imagine that the large amount of gas
lost by stars will expand into the Dark Matter halo and heat up to
an energy overwhelming the gravitational potential, it will escape
the galaxy. Most likely a sort of dynamical equilibrium is reached
in which gas is continuously ejected by stars and lost by the
galaxy. It may happen therefore that some amount of gas is always
present in the galaxy. The question is not trivial because if an
early type galaxy is free of gas and contains only stars, the SED is
expected to drop off long-ward of about $2 \mu m$ and no IR emission
should be detected. However, as already pointed out long ago
by \citet{Guhathakurta86,Knapp89} (see also Fig. \ref{X2}), many
early-type galaxies of the local universe emit in the IR. The origin
of this flux in the MIR/FIR is likely due to dust present
in a diffuse ISM which, heated up by the galactic radiation field,
emits at those wavelengths. Therefore to match the IR emission one
has to allow for some amount of diffuse ISM. An interesting
question to rise is therefore: how much gas can be present today in
an elliptical galaxy and how is it distributed across the galaxy?
Even if we can correctly estimate the amount of gas ejected by
stars, the fate of this gas goes beyond the possibilities of
classical semi-analytical models.

As a third point, note that when we fold many
SSPs to calculate a galaxy SED using the classical EPS technique we
simply convolve their fluxes with the SFH of the galaxy. Many
classical spectrophotometric semi-analytical models of galaxies are
built in this way: there is no dust at the level of SSPs and again
no dust at the level of the galaxy model \citep[see
e.g.][]{Arimoto87, Arimoto90, Bruzual93, Tantalo96, Kodama97, 
Tantalo98, Buzzoni02, Buzzoni05}.
To calculate the emission by dust, a higher level
of sophistication of the model is required. Indeed one has to
develop a model in which the sources of radiation and the
emitting/absorbing medium are distributed, to face and solve the
problem of the radiative transfer simulating in a realistic way the
interactions among the various physical components of a galaxy.
Among recent models of this kind are those by
\citet{Silva98}, \citet{Devriendt99} and \citet{Takagi03}.

\subsection{Improving upon semi-analytical models}

Two drawbacks of the semi-analytical models concern
therefore: (1) the description of galactic wind, which is supposed
to occur within a finite time interval and (2) the star formation
history that is reversed allowing longer SFH for more massive
galaxies. These two problems, combined with a lack of geometrical
information about the distribution of gas and dust, make
semi-analytical models not suitable to calculate properly the IR
emission of early type galaxies. To improve upon them we need to use
the results obtained from dynamical simulations. They have shown to
be able to properly model the ejection of gas by the galaxy as a
sort of continuous process, taking place whenever a gas particle
heated up by various mechanism has acquired a velocity greater than
the escape velocity \citep[see
e.g.][]{Carraro98,Kawata01,Springel01,Chiosi02}. They are able to
reproduce the SF history of early-type galaxies both in the context
of the monolithic collapse scenario \citep{Kawata01,Chiosi02} and
recently in the context of hierarchical scenario \citep{DeLucia06}.

Finally, the galaxy is no more a mass point, but a fully
three-dimensional structure of the galaxy is available with spatial
distribution of stars and gas.

\citet{Merlin06}, with the aid of \textit{N-Body Tree-SPH}
simulations based on quasi-cosmological initial conditions in the
standard-Cold Dark Matter scenario (S-CDM), modelled the formation
and evolution of two early-type galaxies of different total mass
(Dark + Baryonic Matter in the cosmological proportions 9:1). The
total masses under considerations are  $1.62\times 10^{12} M_\odot$
(Model A) and $0.03\times 10^{12} M_\odot$ (Model B). The galaxies
have been followed from their separation from the global expansion
of the universe to their collapse to virialized structures, the
formation of stars and subsequent nearly passive evolution. They are
followed for a long period of time, i.e. 13 Gyr (Model A) and 5 Gyr
(Model B). In any case, well beyond the stages of active star
formation which occurs within the first 3 to 4 Gyr (see below). The
models take into account radiative cooling by several processes,
heating by energy feed back from supernova explosions (both Type I
and II) and chemical enrichment. All the models conform to the
so-called \textit{revised monolithic scheme}, because mergers of
substructures have occurred very early in the galaxy life. Some
parameters and results of the two models are summarized in Table
\ref{tabcosmo}. Note that the shape of the resulting galaxies is
nearly spherical both in Dark Matter and stars.

\begin{table}
\caption{Initial parameters for  the dynamical simulations of
\citet{Merlin06} in the Standard CMD scenario. Masses are in units
of $10^{12}M_\odot$, radii are in kpc and ages are in Gyr.} \centering
\begin{tabular}{|l|l|l|l|}
\hline
Model                                   & A              & B\\
Cosmological background                 & S-CDM          &  S-CDM       \\
\hline
Initial redshift                        & 50             & 53          \\
\hline
$\Omega_m$                              & 1              & 1            \\
\hline
$H_0 = 50 \mbox{ } km Mpc^{-1} s^{-1} $ & 50             & 50           \\
\hline
Gas particles & 13719 & 13904 \\
\hline
CDM particles & 13685 & 13776 \\
\hline
Total Mass   & $ 1.62 $ & $ 0.03 $ \\
\hline
Initial baryonic mass fraction  & 0.10 & 0.10  \\
\hline
Present gas mass &0.062 &0.0004 \\
\hline
Present star mass &0.091 &0.0029 \\
\hline
$M_{star}/M_{baryons}$ & 0.556& 0.82\\
\hline
Initial radius  & 33 & 9\\
\hline
Half-Mass radius of stars & 7 & 1\\
\hline
Half-Mass radius of DM &52 &15\\
\hline
Effective radius of stars &5.2 & 0.8 \\
Present virial radius  &300  & 41 \\
\hline
Axial ratio b/a (stars) &  1.08 & 1.04\\
\hline
Axial ratio c/a (stars) &1.07 & 1.00\\
\hline
Axial ratio b/a (Dark Matter) &  1.14 & 1.14\\
\hline
Axial ratio c/a (Dark Matter) &1.17 & 0.96\\
\hline
Age of the last computed model & 13 & 5\\
\hline
\end{tabular}
\label{tabcosmo}
\end{table}

\begin{figure}
\resizebox{\hsize}{!}{\includegraphics{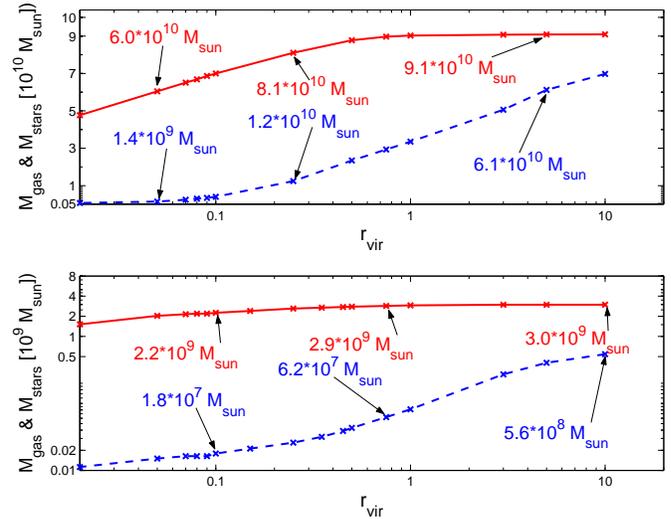}}
\caption{{\it Top panel}: The masses of stars (continuous line) and
gas (dashed line) for the dynamical model of $1.6 \cdot 10^{11}
M_{\odot}$ baryonic mass as a function of the number of virial radii.
The virial radius for this model is about $300$ Kpc. {\it Bottom
panel}: The same of the upper panel, but for the dynamical model of
baryonic mass $3.5 \cdot 10^{9} M_{\odot}$. For this model the
virial radius is about $42$ Kpc.} \label{RaggiViriali}
\end{figure}

The third drawback of classical semi-analytical model was the lack
of the description of the dusty component, that for our purposes
needs to be included. The semi-analytical chemo-spectro-photometric
model developed by \citet{Piovan06b} allows us to overcome this
issue. It takes into account not only the geometrical structure of 
galaxies of different morphological type, but also the effect of dust 
in converting the UV and Optical light in far IR radiation. In brief
the \citet{Piovan06b} model follows the infall scheme, allows for
the onset of galactic winds, and contains three main components: (i)
the diffuse interstellar medium composed of gas and dust whose
emission and extinction properties have been studied in detail by
\citet{Piovan06a}, (ii) the large complexes of molecular clouds in
which new stars are formed and (iii) the stars of any age and
chemical composition. The total gas and star mass provided by the
chemical model are distributed over the whole volume by means of
suitable density profiles, one for each component and depending on
the galaxy type (spheroidal, disk and disk plus bulge). The galaxy
is then splitted in suitable volume elements to each of which the
appropriate amounts of stars, molecular clouds and interstellar
medium are assigned. 
Each elemental volume absorbs radiation from all other volumes 
and from the interstellar medium in between. The elemental volume also 
re-emits the absorbed light and produces radiation by the stars that 
it contains. 
On the other hand, the star formation, the initial mass function, the 
chemical enrichment of the \citet{Piovan06b} model are much similar to those 
by \citet{Bressan94, Tantalo96, Tantalo98, Portinari98}.

\subsection{Coupling dynamical simulations and dusty population synthesis models}

\begin{figure}
\resizebox{\hsize}{!}{\includegraphics{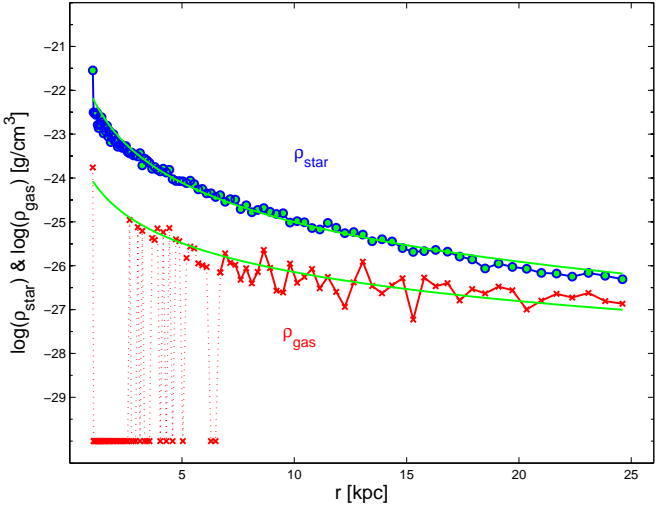}} \caption{The
averaged density profiles of gas and stars for the model of $1.6
\cdot 10^{11} M_{\odot}$ baryonic mass at the age of $13$ Gyr. The
continuous lines are the best fit obtained.} \label{Fit_RHO}
\end{figure}

The description of an early-type galaxy as far as predicting
its spectro-photometric infrared properties can be therefore
realized with a suitable combination of dynamical and
spectro-photometric approaches. Coupling the dynamical models with
spectro-photometric synthesis requires a number of steps that
deserve some remarks.

\subsubsection{Radial density profiles.} 
Fig. \ref{RaggiViriali} shows the
cumulative distribution of gas and stars as a function of the radial
galactocentric distance normalized to the virial radius for model A
(top panel) and model B (bottom panel). The gas is generally
distributed in the external regions of the galaxy and steeply
decreases inward. In contrast the stars are more concentrated toward
the centre. The gradients in the spherically averaged star- and gas-
content provided by the dynamical models are the primary information
to load into the spectro-photometric code of \citet{Piovan06b}. They
allow us to infer the amount of gas contained within a given radius
or within a given aperture. We fix the total dimension of portion of
the average model producing the IR flux at a diameter $D_{gal} = 25$
kpc, consistent with the mean galaxy size of the observed sample.

As the spectro-photometric code of \citet{Piovan06b} suited to
describe early-type galaxies is written in spherical symmetry, we
have to derive suitable spherical distributions for the density of
stars and gas to be used into the model. The task is facilitated by
the nearly spherical shape of the dynamical models. To this aim, we
consider the sphere of radius $R_{gal}$ centred at the centre of
mass of the stellar component. The sphere is then divided in a
number of thin spherical shells whose derived average density of
stars and gas is shown in Fig. \ref{Fit_RHO}. Even if the
centre of mass of the star and gas distributions may not be exactly
coincident, this not relevant here, so that the same coordinate centre 
can be used for both components.

In order to secure a smooth behaviour at the galaxy radius $R_{gal}$
the star and gas density profiles are represented by the law:

\begin{equation}
\rho _{i}=\rho _{0i}\left[ 1+\left( \frac{r}{r_{c}^{i}}\right)
^{2}\right] ^{-\gamma _{i}}  \label{rhostar_ell}
\end{equation}

\noindent where  $``i"$  stands for $``stars"$ or $``gas "$,
$r_{c}^{i}$  are the corresponding core radii. The above
representation is more suited to our aims than the classical King
law. The fits are shown in Fig. \ref{Fit_RHO} (solid lines). They
are normalized in such a way that the integral over the galaxy
volume corresponds to the amount of gas contained inside $R_{gal}$.

\subsubsection{Star formation rate.} 
In the dynamical models, the period of
intense star formation, during which most of the star mass is built
up, is confined within the first 3 to 4 Gyr. In Model A this is
followed by a long tail of minimal stellar activity which continues
forever. If this activity would be real, we would expect a background of
young stars giving rise to a significant emission in the UV-optical
region up to the present, which is not compatible with the observed
spectra of typical early-type galaxies. As already pointed out by
\citet{Merlin06} this minimal stellar activity is an artefact of the
poor mass-resolution for the baryonic component, in other words the
low number of particles considered in the numerical simulations. To
cope with this, we simply set to zero the star formation rate when
only one or two star particles are involved. This is equivalent to
cut the star formation rate for ages older than about 5 Gyr. The
problem does not occur with model B simply because the last computed
models is at 6 Gyr.

\begin{figure}
\resizebox{\hsize}{!}{\includegraphics{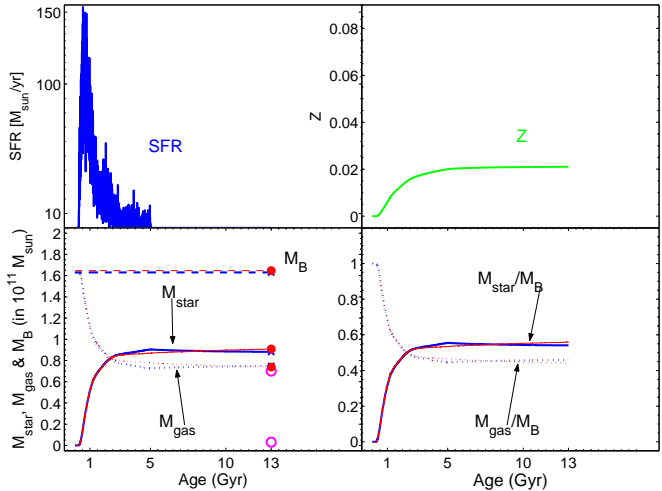}}
\caption{Basic quantities of the chemical models for a prototype
early-type galaxy of $1.6 \cdot 10^{11} M_{\odot}$ baryonic mass as
function of the age. {\it Top-left panel}: The star formation rate
obtained from the chemo-dynamical model of $1.6 \cdot 10^{11}
M_{\odot}$ baryonic mass (SFR). {\it Top-right panel}: The evolution
of the metallicity $Z$. {\it Bottom-left panel}: The masses for both
the classical and the chemo-dynamical model of baryonic matter
(thick and thin dashed lines), stars (thick and thin continuous
lines) and gas (thick and thin dotted lines). The mass of gas and
stars inside $25$ Kpc is also shown (open circles) for the age of
$13$ Gyr. {\it Bottom-right panel}: The fractional masses for both
the models of stars (thick and thin continuous lines) and gas (thick
and thin dotted lines).} \label{BIG_SF}
\end{figure}

\begin{figure}
\resizebox{\hsize}{!}{\includegraphics{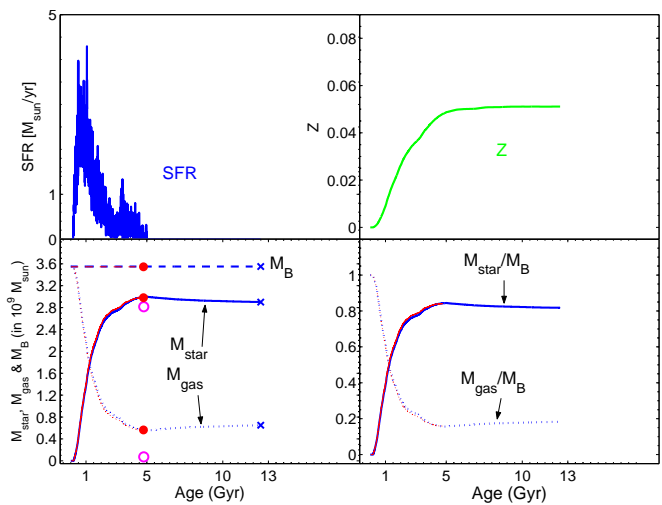}}
\caption{The same as in Fig. \ref{BIG_SF} but for the model with
total baryonic mass of $3.5 \cdot 10^{9} M_{\odot}$.}
\label{SMALL_SF}
\end{figure}

\subsubsection{Checking dynamical models against chemo-spectro-photometric
models.} 
To this aim we plug the star formation history (SFH) of
dynamical models into the chemical code of \citet{Portinari98}.
The closed-box approximation is adopted. The total
baryonic mass of the chemical models is the same as in the dynamical
ones. Equally for the initial mass function of the stars composing
each star particle: \citet{Kroupa98} in our case. In Fig.
\ref{BIG_SF} we show the results obtained by inserting the SFH of
Model A into a classical chemical model with total baryonic mass
$M_B$ equal to $1.6 \cdot 10^{11} M_{\odot}$. The top panels display
the adopted SFH (left) and the gas metallicity of the chemical
model, respectively. The bottom left panel shows the temporal
variation of the star mass $M_{star}$ and gas mass $M_{gas}$,
whereas the bottom right panel shows the ratios $M_{star}/M_B$ and
$M_{gas}/M_B$  for both the dynamical (thin lines) and chemical
model (thick lines). The agreement is very good thus confirming the
internal consistency between the descriptions of the same object. We
also show the amount of gas at $13$ Gyr contained in the whole
galaxy for both the dynamical (heavy dots) and the classical
chemical models (open circles) and the amount of gas contained
inside $R_{gal}$ (open squares). Indeed there is little gas left
over inside the 25 kpc radius region. Similarly in Fig.
\ref{SMALL_SF}, we show the results obtained inserting the SFH of
Model B into a classical chemical model with total baryonic mass of
$3.5 \cdot 10^{9} M_{\odot}$. The only difference is that the
maximum age of the dynamical model is 5 Gyr. This cross
checking of the models is particularly significant because: first,
it secures that the results of the analytical models fairly
reproduce those of the dynamical simulations as far as some
important features are concerned; second, it secures that we can
safely use the result of chemical models to prolong the evolutionary
history of Model B up to the present; third, that we can safely
apply the population synthesis technique of \citet{Piovan06b}.

Knowing the amount of gas, we need to specify the fraction of it in
form of dust to finally be able to derive the whole SED from X to
FIR and look for relationships between the luminosity in the X, B
and FIR pass-bands we want to interpret. Our models, both
semi-analytical and chemo-dynamical, are not suitable to describe
the evolution of the compositions and abundances of \textit{both}
gas and dust phases. 
The relative proportions of the various
components of the dust would require the detailed study of the
evolution of the dusty environment and the complete information on
the dust yields, as in the models of \citet{Dwek98,Dwek05}. 
This
would lead to a better and more physically sounded correlation
between the composition of dust and the star formation and chemical
enrichment history of the galaxy itself, however at the price of
increasing the complexity and the uncertainty of the problem.

The key parameter to calculate the amount of dust is the dust-to-gas
ratio, defined as $\delta =M_{d}/ M_{H}$, where $M_d$ and $M_{H}$
are the total dust and hydrogen mass, respectively. For the Milky
Way and the galaxies of the Local Group, $\delta$ is estimated to
vary from about $1/100$ to $1/500$ and typical values $\delta =
0.01$, $\delta =0.00288$ and $\delta =0.00184$ are used for the
Milky Way (MW) and the Large and Small Magellanic Clouds (LMC and
SMC). These dust-to-gas mass ratios describe a decreasing sequence,
going from the MW to the LMC and SMC. Since these galaxies also
describe a sequence of decreasing metallicity, a simple assumption
is to hypothesize $\delta \varpropto Z$ in such a way to match the
approximate results for MW, LMC and SMC: $\delta
=\delta_{\odot}\left(Z/Z_{\odot}\right)$. \noindent This relation
simply implies that the higher is the metal content of a galaxy, the
higher is the abundance of grains per $H$ atom. However, the
metallicity difference does not only imply a difference in the
absolute abundance of heavy elements in the dust, but also a
difference in the composition pattern as a function of the
star formation history \citet{Dwek98,Dwek05}. Despite these
uncertainties \citep{Devriendt99}, the relation $\delta \varpropto
Z$ is often adopted to evaluate the amount of dust in galaxy models
\citep[e.g.][]{Silva98} by simply scaling the dust content adopted
for the ISM of the MW to the metallicity under consideration.

The $1.6 \cdot 10^{11}M_{\odot}$ and $3.5 \cdot 10^{9}
M_{\odot}$ galaxy models reach an average metallicity of solar and
slightly more than twice solar, respectively. To describe them we
have adopted the description of \citet{Piovan06a, Piovan06b} where a
model of dusty ISM taking into account different metallicities is
built. The problem however remained unsettled for metallicities
higher than the solar one, where relative proportions holding good
for the MW average diffuse ISM model have been adopted and the
amount of dust scaled with $\delta \varpropto Z$. Therefore, for the
$1.6 \cdot 10^{11}M_{\odot}$ galaxy with solar metallicity the MW
diffuse ISM model has been adopted $\left(\delta =\delta_{\odot}
\right)$, while for the $3.5 \cdot 10^{9} M_{\odot}$ model we
followed the $\delta \varpropto Z$ relation, using the MW average
pattern of dust composition.

The connection between the results of this model and the observed
diagrams are discussed in the following section.

\section{Discussion} \label{Discussion}

Our data confirm and extend the previous relations existing between
various tracers of the ISM in galaxies of different morphological
types.

In the literature the relation found by \citet{bregman} between
S$_{CO}$ and S$_{100}$ indicates a direct proportionality (slope=1)
between the two fluxes and differs from that of \citet{solomon},
that exhibits a steeper gradient. Our relation (\ref{eqCO_100})
agrees quite well with the proportionality found by \citet{bregman},
the slope we found being equal to 1.06. The similarity between the
two curves in Figure \ref{CO_100} is evident. This link derives, as
described in the introduction, from the excitation of gas clouds by
the currently forming stars and by the warming of the dust present
in the galaxy.

\subsection{Late-type galaxies}

In late-type galaxies (t$>$Sb) our data show the existence of a
linear relation between soft X-ray fluxes and other indicators of
recent and current star formation, such as the B and FIR luminosity
respectively (equations \ref{XBlate} and \ref{XFIR}). This is known
since the first X-ray observations of large samples
\citet{fabbiano0} and this connection between B and X-ray luminosity
in late type galaxies has been interpreted as due to the
contribution of discrete X-ray sources, whose number is proportional
to the quantity of already formed stars \citep{ciotti, beuing}. The 
recent work of \citet{fabbiano}, that is able to resolve the single
X-ray binaries in 14 galaxies, indicates that the X-ray luminosity
produced by discrete sources is related to B luminosity by a similar
relation, with an intercept value of -3.63, similar to our -3.85 of
equation \ref{XBlate}.

In addition to the interstellar radiation, that is proportional to
the number of already formed stars, the X-ray emission is produced
also by HII regions, where there is an ongoing vigorous star
formation \citep{david}. This latter contribution appears more
evident in FIR light and may explain the existence of a similar
linear relation between L$_X$ and L$_{FIR}$.

\subsection{Early-type galaxies}
In early type galaxies the behaviour of these relations is
quite different. For most of these galaxies, the star
formation is exhausted and it may be present in a few of them,
eventually fed by gas accretion phenomena. Different mechanisms have
been suggested to explain the X-ray emission in this kind of
galaxies. In particular the main ones are the thermal emission due
to hot ISM and the emission generated by a relatively old population
of end objects of stellar evolution, composed by Type I supernovae
remnants and low-mass X-ray binaries not yet evolved. In particular,
for the fainter galaxies the X-ray emission is compatible with
discrete sources and seems to be dominated by compact accreting
systems, while for the brighter objects the emission from hot
diffuse gas still present in the galactic potential well is present
as additional component \citep{beuing}. The number size of this
population of relatively old objects is well represented by the
total blue luminosity of the galaxy. For this reason the X-ray
fluxes are still linked in early type galaxies with the total blue
luminosity, representing the more recent part of the history of star
formation in the galaxy.

\begin{figure}
\resizebox{\hsize}{!}{\includegraphics{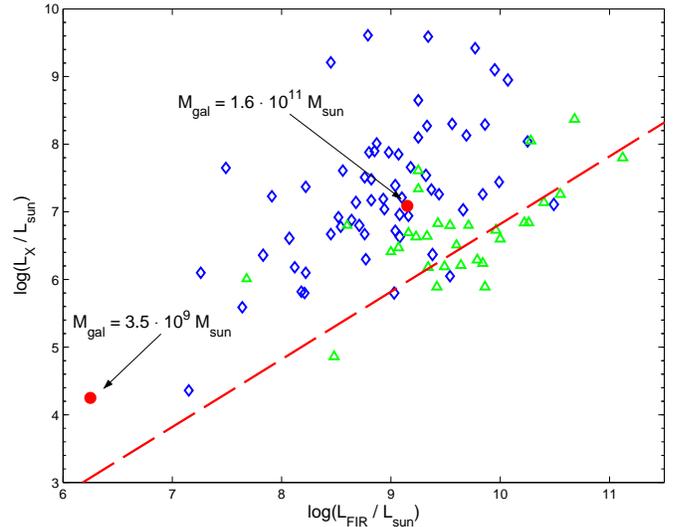}}
\caption{$L_{FIR}$ vs. $L_{X}$ diagram as in Figure 3, right panel,
but with the values deduced for the two template models of
early-type galaxy (filled circles). The diamonds represents the
$E-S0$ galaxies, while the triangles the $S0a-Sb$ galaxies. The
masses of the two theoretical models are also indicated.}
\label{LxLfirES}
\end{figure}

In the FIR however, since the star formation in most of
these systems is almost exhausted, mechanisms different from the
emission from warm dust heated by the newly born stars predominate.
The FIR emission comes from circumstellar dusty shells around AGB
stars and from an interstellar medium due to the outflow of dusty
gas from AGB and RGB stars, as it has been described in Sect.
\ref{Modelling}.

The key point to interpret the observed trends is that we
deal with an emission coming from a more or less small amount of
dust distributed over all the galaxy and heated by an average
interstellar radiation field due to all the stars of any age. The
situation is quite different from what happens for instance in
starburst galaxies where high optical depth dusty regions reprocess
the light coming from newly born stars embedded in the parental
environment. We can therefore conclude that in most of our
early-type galaxies the mechanism of IR emission is not strictly
related to the star formation and the link between the younger
generations of stars and dust emission is lost. For these reasons
one may expect that the soft X-ray luminosity in early type galaxies
is traced by the total blue luminosity {\it but not} by the FIR
luminosity. With the end of the star formation, the far infrared
emission of these galaxies has faded out and an early type galaxy
with the same $L_X$ of a late type will have a lower $L_{FIR}$. This
could explain the location of the points in Fig. \ref{X2} (right
panel), on the left side of the linear relation.

To check if this interpretation is correct we try to apply the
detailed chemo-dynamical spectrophotometric model described in the
previous section, in such a way to estimate the luminosities
produced by the stars in connection with the various phenomena
present inside the galaxy, taking into account the contribution by
dust as well. Since the theoretical model can not derive the $L_{X}$
luminosity, we proceed in the following way.

The luminosities $L_{B}$ and $L_{FIR}$ are directly derived
from the model. Then, we assume that the X-ray production of these
galaxies is proportional to $L_{B}$ according to our relation
(\ref{XBearly}). In this way we may estimate the expected X-ray flux
and define a representative point in the $L_{FIR}$ vs $L_{X}$ plot.

We start considering two template models, in which all the
parameters are fixed using the clues coming from the dynamical
simulations of \citet{Merlin06}, as described in Sect.
\ref{Modelling}. The King profiles represented in Fig. \ref{Fit_RHO}
are similar for all the components, with $\gamma_{stars} \simeq
\gamma_{gas} \simeq 1.5$ and $r_{c}^{stars} \simeq r_{c}^{gas}
\simeq 0.5$ Kpc, while the dimension of the galaxy is an average one
corresponding to most of the galaxies available in the catalogue.
The SFH is exactly the one obtained by the dynamical simulations.
The two values of $L_{FIR}$ and $L_{X}$ obtained for the $3.5 \cdot
10^{9} M_{\odot}$ and $1.6 \cdot 10^{11} M_{\odot}$ baryonic mass
models are plotted in Fig. \ref{LxLfirES}. The more massive galaxy
fits well into the region defined by the really observed galaxies,
while we can notice as the model of smaller mass, even if falling up
to the linear relationship as we could expect, belongs to a region
not covered by the observed data. The calculated levels of emission
$L_{X}$ and $L_{FIR}$ of this galaxy are very low and for this
reason they belong to a region where we do not have enough
observations. The weak $L_{FIR}$ emission of this galaxy can be
explained by the dynamical evolution in which almost all the gas is
consumed to form stars and the galactic winds are very efficient
\citep[see][ for more details about galactic wind in low mass
galaxies]{Chiosi02}. Therefore, even if the trend of this galaxy is
the expected one for early-type galaxies (the model stays above the
linear relation), nothing safer can be said, because we lack
observed data in that region of the diagram.

\begin{figure}
\resizebox{\hsize}{!}{\includegraphics{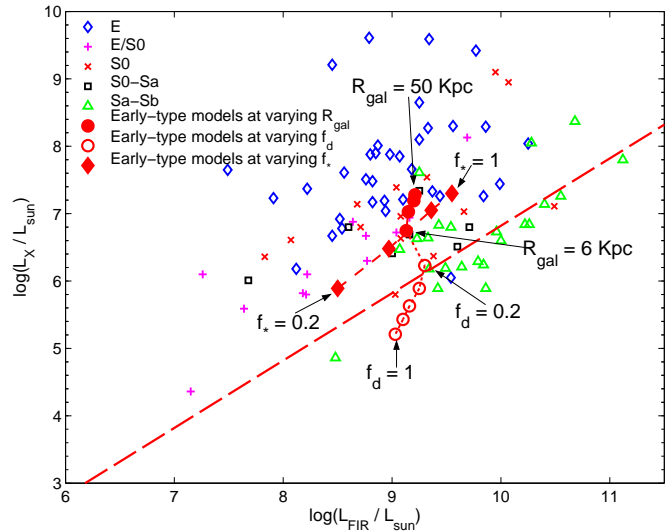}} \caption{Models
of the $1.6 \cdot 10^{11} M_{\odot}$ baryonic mass galaxy in the
$L_{FIR}$ vs. $L_{X}$ diagram at varying the galactic radius (filled
circles), the mass of stars (filled diamonds) and, finally, the mass
of gas (empty circles). We represented: $E$ galaxies (empty
diamonds), $E/S0$ galaxies (pluses), $S0$ galaxies (crosses), $S0s$
galaxies (empty squares), $Sa-Sb$ (empty triangles).} \label{Raggi}
\end{figure}

Much more interesting is the model of higher mass. 
The calculated
luminosities of the model, with its exhaustion of the star
formation, seem to agree well with the observations of early-type
galaxies.
However, the model needs to be checked against other
possibilities, at the purpose to understand the way in which the
various parameters of the model influence the spreading of
early-type galaxies into the observational data. First of all we
have to check the effect of the geometrical parameters and of the
masses of stars/gas.

\subsubsection{The galactic radius}
In fig. \ref{Raggi} we show the model of $1.6 \cdot 10^{11}
M_{\odot}$ baryonic mass at varying the galactic radius, keeping the
galactic center in the center of mass of the stellar component. The
radii taken into account range from $6$ Kpc to $50$ kpc. All the
other parameters are fixed. Four models are represented (filled
circles) and connected by a continuous line and the smaller and
bigger models are marked using an arrow. For larger radii we observe
an increase of both $L_{FIR}$ and $L_{X}$, with a more emphasized
increase in $L_{X}$. Since the density profile is unchanged, both
the increases in luminosity are simply due to the bigger amount of
material considered taking into account larger radii in the
dynamical simulation. The stronger increase in $L_{X}$ than
$L_{FIR}$ can be simply explained. $L_{X}$ is linearly related to
$L_{B}$, that is directly connected to the stellar luminosity. The
stellar component is more massive and more concentrated toward the
centre than the gaseous one (Fig. \ref{RaggiViriali}, upper panel).
It follows that at increasing radius we introduce into the models
more stars and more gas, but the added amount of stars is bigger
than the gaseous one, shifting $L_{B}$ (and then the linearly
related $L_{X}$) more than $L_{FIR}$. Finally, we observe how, even
taking into account the smallest radius of $6$ Kpc, it is not
possible to move the theoretical point near the linear relation
holding for spirals.

\subsubsection{The masses of stars and gas}
We also investigated in Fig. \ref{Raggi} what happens if we forget
about the clues coming from dynamical simulations on the masses of
stars and gas and we arbitrarily start varying the amounts of stars
or gas, keeping all fixed. Filled diamonds represent the shift of
the model of lowest radius if we are changing the mass of stars
inside $R_{gal}$, going in fraction from $f_{*}=0.2$ to $f_{*}=1.0$,
with respect to the total amount of stars in the dynamical model.
The effect is simply to move the point along a line about parallel
to the linear relation. A smaller amount of stars imply directly a
lower luminosity $L_{B}$ (and therefore a lower $L_{X}$), but also a
lower $L_{FIR}$, because the weaker radiation field makes dust
cooler and shifts the peak of dust emission to wavelengths longer
than $100 \mu m$, with the result of a smaller $L_{FIR}$. Finally,
with open circles we show in Fig. \ref{Raggi} five models obtained
at fixed amount of stars and at varying the mass of gas (and
therefore of dust) from $f_{d}=0.2$ to $f_{d}=1.0$, in fraction
respect to the total amount of gas in the dynamical model. The
effect of this huge increase of the mass of diffuse gas and dust (in
the original model at $R_{gal} = 6 $Kpc only $0.03 \%$ of the gas is
inside $R_{gal}$) is to shift the models straight toward the linear
relation. It can be explained in the following way. Increasing the
amount of diffuse gas/dust (with all the parameters fixed and the
star formation exhausted) implies more absorption of the stellar
radiation and therefore a smaller $L_{B}$ (and $L_{X}$). On the
other side, $L_{FIR}$ remains almost unchanged or becomes smaller.
The reason is that the strongly increased mass of dust makes the
average stellar radiation field weaker and therefore the increased
emission of dust (due to the bigger mass) peaks at wavelengths
longer than $100 \mu m$, leaving $L_{FIR}$ almost unchanged. Even if
in this way we can shift the model toward the linear relation, the
situation is physically unrealistic, requiring a huge amount of
gas/dust concentrated in the centre of an early-type galaxy with
exhausted star formation, which is not commonly observed and also
not predicted by dynamical models.

\begin{figure}
\resizebox{\hsize}{!}{\includegraphics{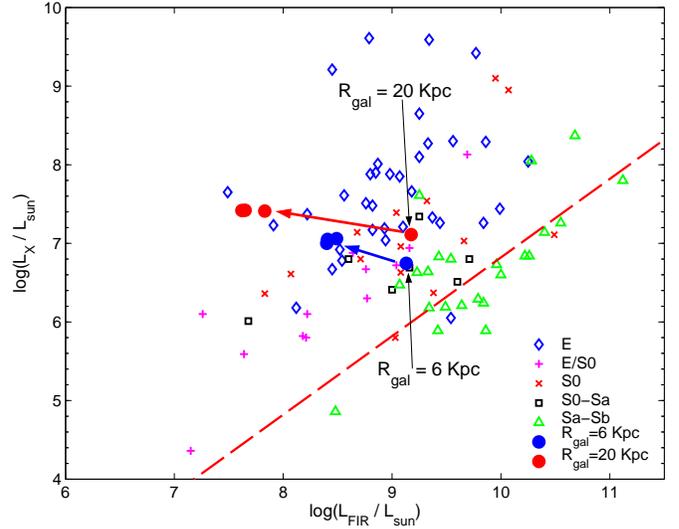}} \caption{Models
of the more massive galaxy of $1.6 \cdot 10^{11} M_{\odot}$ baryonic
mass in the $L_{FIR}$ vs. $L_{X}$ diagram for various core radii
$r_{c}^{stars}$ and $r_{c}^{gas}$, plotted as filled circles. The
thin arrows indicate the original model, while the thick arrows
indicate the shift of the model for various core radii. We
represented: $E$ galaxies (empty diamonds), $E/S0$ galaxies
(pluses), $S0$ galaxies (crosses), $S0s$ galaxies (empty squares),
$Sa-Sb$ (empty triangles).} \label{Profili}
\end{figure}

\subsubsection{The scale radii}
Further geometrical parameters that must be examined are the
scale radii $r_{c}^{i}$ of the King's laws - eqn.
(\ref{rhostar_ell}) - that describe the distribution of the stellar
and gaseous components. The averaged profiles showed in Fig.
\ref{Fit_RHO} and used for the models of Figs. \ref{LxLfirES} and
\ref{Raggi} are both characterized by $r_{c}^{i} \simeq 0.5$,
allowing for a concentrated amount of stars and gas in the inner
regions. Keeping all the other parameters fixed, we investigated
what happens if we allow for a uniform distribution of one or both
the physical components. Three cases have been considered: a uniform
distribution of gas keeping fixed the stellar one $\left(r_{c}^{gas}
\rightarrow \infty, r_{c}^{stars} \simeq 0.5 \right)$, a uniform
distribution of stars keeping fixed the gaseous one
$\left(r_{c}^{stars} \rightarrow \infty, r_{c}^{gas} \simeq 0.5
\right)$ and, finally, a uniform distribution of both the components
$\left(r_{c}^{stars} \rightarrow \infty, r_{c}^{gas} \simeq \infty
\right)$. The results are shown in Fig. \ref{Profili}, for two radii
of the galaxy model, $R_{gal}=6$ Kpc and $R_{gal}=20$ Kpc,
respectively. The three different distributions give a similar
result: a weaker $L_{FIR}$, shifting the point to the left, and a
slightly higher $L_{X}$.

This can be explained in the following way: for
$r_{c}^{stars}$ and $r_{c}^{gas}$ both $\simeq 0.5$, the diffuse ISM
and the stars are both concentrated in the inner region of the
galaxy with a density of stars/gas of many order of magnitude bigger
than the outer regions. This is the best condition to produce high
$L_{FIR}$, because we have that the regions of higher density of
dust are the same in which there is also the higher average
radiation field heating dust. The spatial distribution of the ISM
favors the interaction with the stellar radiation. When we destroy
this coupling between stellar emission and density of gas, as we do
allowing for a uniform distribution of gas or stars or both, the
emission in the $L_{FIR}$ becomes weaker. The weakening of the dusty
emission is stronger for the bigger radius of $20$ Kpc because in
all the three cases one or both the components are distributed over
a huge galactic volume and we have low density of gas eventually
coupled with weak radiation field. For the $6$ Kpc model, even if
the coupling in the central regions is destroyed, the galaxy is
small enough to keep a good level of $L_{FIR}$, even when the matter
is equally distributed across all the galaxy volume.

\subsubsection{The star formation history}

Last and main point to be examined is how varying the star
formation history affects the position of the galaxies into the
$L_{FIR}$ vs $L_{X}$ plot. In Figs. \ref{Raggi} and \ref{Profili}
the galaxies of different morphological type form a sequence that,
going from systems in which the star formation got exhausted long ago to
systems in which star formation is still active, moves toward the
linear relation and suggests the key role played by the star
formation. First of all we calculate the $L_{FIR}$ and $L_{X}$
obtained by the SEDs and the models by \citet{Piovan06b} of real
galaxies of the local universe: three
spiral galaxies $(M100, M51$ and $NGC6946)$ and two starburst
galaxies $(Arp220$ and $M82)$. The key point is that the SFHs of
these galaxies allow us to cover a good number of different star
formation histories. All these SFHs, unlike the ones of the ellipticals
obtained by dynamical simulations, never end and in the case of the
two starbursters a strong burst of star formation is added in the
last millions of years. A huge amount of $L_{FIR}$ comes therefore
from the young and deeply obscured region of star formation and not
only from the diffuse component.

\begin{figure}
\resizebox{\hsize}{!}{\includegraphics{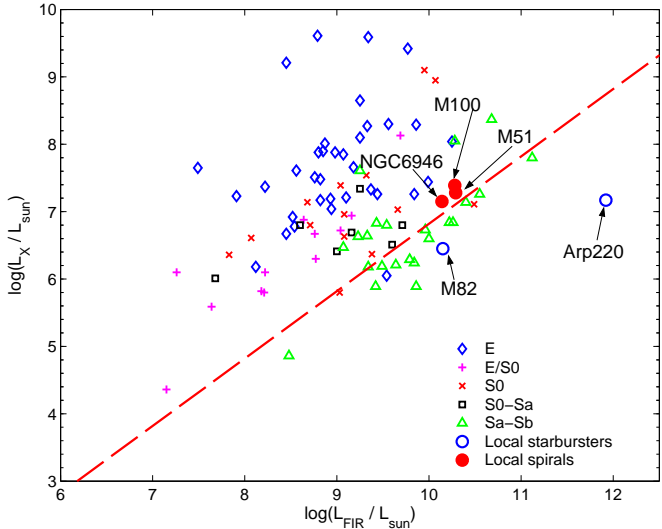}}
\caption{$L_{FIR}$ vs $L_{X}$ for galaxy models of spirals and
starbursters. Five theoretical models, taken from \citet{Piovan06b},
are represented: three spirals (M100, M51 and NGC6946) and two
starbursters (Arp220 and M82). Different morphological types are
plotted as in Fig. \ref{Profili}.} \label{RealGalaxies}
\end{figure}

\begin{figure}
\resizebox{\hsize}{!}{\includegraphics{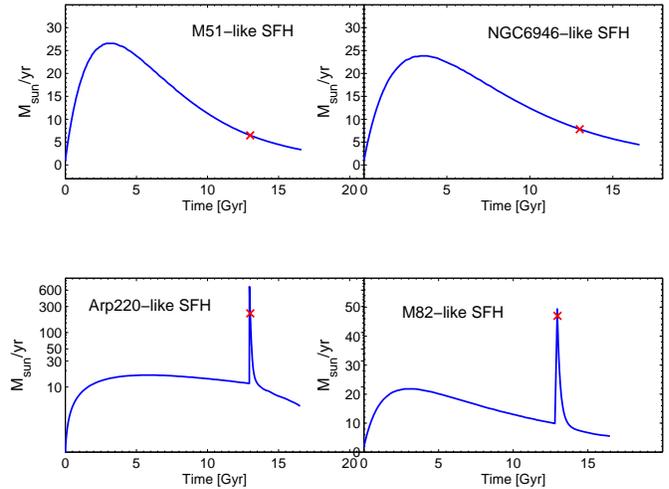}} \caption{Different
adopted star formation histories for the model of $1.6 \cdot 10^{11}
M_{\odot}$ baryonic mass. The SFHs of \citet{Piovan06b} have been
re-calculated for the same mass of the dynamical model. The upper
panels represent two spiral-like SFHs, while bottom panels show two
starburst-like SFHs, peaked in the last million years. } \label{SFH}
\end{figure}

The results, presented in Fig. \ref{RealGalaxies}, show that
the three models of spirals stay near the linear relation, while the
two starbursters stay below the line, with the model of $Arp220$,
powered by a huge burst of star formation falling well below the
linear relation. The stronger is the emission coming from the
regions of star formation and the bigger is the shift toward higher
$L_{FIR}$ and lower $L_{X}$ (due to the lower $L_{B}$). The results
obtained from the models are quite similar to the observational data:
for $M100$ we get $(L_{FIR},L_{X})=(10.28,7.29)$ with the observations giving
$(10.37,7.01)$, for $Arp220$ we have $(L_{FIR},L_{X})=(11.92,7.17)$
compared with $(11.99,7.60)$ and for $M82$ we get
$(L_{FIR},L_{X})=(10.15,6.45)$ against $(9.79,6.31)$.
However, these galaxy models, even if they well represent
real galaxies, differ in many parameters from the early-type galaxy
model of $1.6 \cdot 10^{11} M_{\odot}$, like geometry and mass.
These parameters, together with the SFH, obviously concur to
determine the position of the models into the $L_{FIR}$ vs $L_{X}$
plot. To isolate the effect of the SFH, we first re-calculated the SFHs
of the above five theoretical models, rescaled to
the mass of $1.6 \cdot 10^{11} M_{\odot}$ of the early-type galaxy
model. In Fig. \ref{SFH} we can see four of the five SFH obtained.
Second, we fixed all the geometrical parameters to the same
values used for the average model of $1.6 \cdot 10^{11} M_{\odot}$
early type galaxy. The additional parameters, that is the escaping
time of young stars from parental molecular clouds, the library of
SEDs of young dusty regions and the mass of gas in the diffuse and
molecular component, are fixed to the values used in
\citet{Piovan06b} for spirals and starbursters as appropriate.

\begin{figure}
\resizebox{\hsize}{!}{\includegraphics{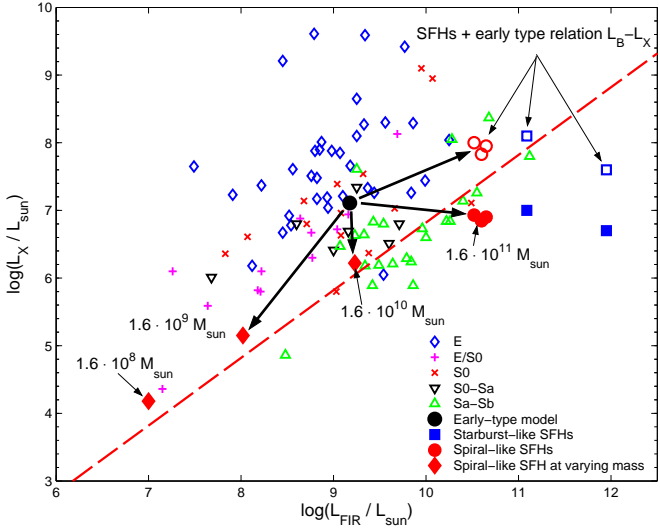}}
\caption{$L_{FIR}$ vs $L_{X}$ for galaxy models of fixed $1.6 \cdot
10^{11} M_{\odot}$ baryonic mass for various SFHs (large circles and
squares) or at fixed SFH for various masses from $1.6 \cdot 10^{8}
M_{\odot}$ to $1.6 \cdot 10^{11} M_{\odot}$ (large filled diamonds).
Large circles and squares represent respectively models with
spiral-like SFH and starburst-like SFH. Large filled circles and
squares are obtained using eqn. \ref{XBlate}, while large open
circles and squares stand upon eqn. \ref{XBearly}. Different
morphological types are plotted as in Fig. \ref{Profili}.}
\label{varyingSFHs}
\end{figure}

In Fig. \ref{varyingSFHs} we finally show the results
obtained as a function of the SFH of the galaxy of $1.6 \cdot 10^{11}
M_{\odot}$, keeping all the other parameters fixed. It is
interesting to observe that since now the star formation never ends
and the galactic wind is not included, the classical semi-analytical
chemical evolution can be much more safely coupled to the
spectro-photometric code. The effect of varying the SFH at fixed
mass is to enhance the $L_{FIR}$, keeping almost fixed the $L_{X}$
and shifting the points toward the linear relation at higher
infrared luminosities. This is ultimately due to the strong and
efficient reprocessing of the light coming from very young stars,
occurring into the dusty star-forming regions. As a consequence of
this, models with starburst-like SFHs shift, as expected, toward
higher $L_{FIR}$ luminosity than models with spiral-like SFH,
because of the stronger star formation and therefore emission coming
from young dusty regions. This can be also understood if we look in
detail at the relative contribution to $L_{FIR}$ coming from the
regions of star formation (let us define it $f_{MCs})$ and represent
it as usual in $\log(L_{FIR}/L_{\odot}$) and from the diffuse interstellar
medium $(f_{ISM})$. We get the following values: ($f_{SFR}=10.15,
f_{ISM}=10.48$), ($f_{SFR}=10.02, f_{ISM}=10.45$), ($f_{SFR}=9.98,
f_{ISM}=10.37$) for the three models with spiral-like SFHs, while we
have ($f_{SFR}=10.97, f_{ISM}=10.47$) and ($f_{SFR}=11.93,
f_{ISM}=10.56$) for the models with starburst-like SFHs. The stronger
is the contribution from star forming regions, the higher is
$L_{FIR}$ keeping $L_{B}$ (and $L_{X}$) almost unchanged. Models
slightly dominated by the ISM contribution, but with a significant
contribution coming from obscured newly born stars are more suitable
to agree with the linear relation of spirals.

It is worth noticing that in Fig. \ref{varyingSFHs} we show both the
results obtained applying the early-type linear relation between
$L_{X}$ and $L_{B}$ - eqn. (\ref{XBearly}) - and the late-type one -
eqn. (\ref{XBlate}). Since, however, the SFHs used (see Fig.
\ref{SFH}) are typical of late type galaxies (or starbursters), it's
more physically sounded to apply eqn. (\ref{XBlate}) to obtain the
$L_{X}$ luminosity. As last point we calculated also a sequence of
models in which one of the SFHs of the spirals has been chosen
(namely the one of NGC$6946$) with all the parameters fixed and only
the mass is varied. As we see from Fig \ref{varyingSFHs} the effect
of varying the mass is to shift the object in diagonal almost along
the relation. This is simply explained by the smaller amounts of
stars/gas emitting radiation.

\section{Conclusions}

We have been able to describe the relations existing in a galaxy
between the various tracers of the ISM and to fix the coefficients
of the relations existing between FIR, B and X-ray luminosity, both
for early-type and late-type galaxies.

The large set of data we used allowed us to redefine more clearly the 
relation existing between the CO and the 100 $\mu$m fluxes. We found 
that the relation, first obtained by \citet{bregman} for early type 
galaxies, is valid also for late type galaxies. In these galaxies, the 
X-ray flux appears linked also to B and FIR emissions.
 
The only relation lacking from observations, i.e. the one
between L$_X$ and L$_{FIR}$ has been studied by the use of the most
recent chemo-dynamical models coupled with dusty evolutionary
population synthesis.

The calculated luminosities of the models seem to confirm our
hypothesis about a connection between the exhaustion of the star
formation and the "migration" of the early type galaxies above the
linear relation in the L$_X$ vs L$_{FIR}$ plot. In the frame of our
assumptions, we may therefore conclude that the prediction of our
dusty chemo-dynamical models of galaxy evolution is consistent with
the observed lack of a direct relation between $L_{X}$ and $L_{FIR}$
for early type galaxies and is due to the different mechanisms of
production of FIR light in galaxies where the active star formation
is no longer active. In most of our early-type galaxies the
mechanism of IR emission is no more strictly related to the ongoing
star formation and to the reprocessing of the radiation in the dense
regions where new stars are born. The FIR emission comes therefore
most likely from circumstellar dusty shells around AGB stars and
from an interstellar diffuse medium due to the outflow of dusty gas
from AGB and RGB stars. 

Finally, we can summarize that: (i) the SFH
of the galaxies seems therefore to have the stronger effect on the
position of early-type galaxies in the L$_X$ vs L$_{FIR}$ plot; (ii)
other parameters, like the radius of the galaxy and the scale radii
of stars and gas, play a secondary role, even if they can
significantly contribute to the scatter of the models in the region
above the linear relation; (iii) the mass is the main parameter
explaining the scatter of the points along the linear relation.

\begin{acknowledgements}
This research has been partially funded by the University of Padua
with Funds ex 60\% 2005. We acknowledge Prof. C. Chiosi for
useful discussions on theoretical subjects of this paper. L. Piovan is pleased
to acknowledge the hospitality and stimulating environment provided by
Max-Planck-Institut f\"ur Astrophysik in Garching where part of the
work described in this paper has been made during his visit as EARA
fellow on leave from the Department of Astronomy of the Padua
University. We also thank the Referee for the detailed and useful comments
about this topic.\end{acknowledgements}

\end{document}